\documentclass[aps,prl,twocolumn,amsmath,amssymb,floatfix,showkeys]{revtex4-1}
\usepackage[dvipsnames]{xcolor}
\usepackage{graphicx}
\usepackage{amsmath}
\usepackage{amssymb}
\usepackage[colorlinks=true,linkcolor=blue,citecolor=blue,urlcolor=blue]{hyperref}

\makeatletter
\let\frontmatter@footnote@produce\frontmatter@footnote@produce@footnote
\makeatother

\newcommand{\prlsec}[1]{\textit{{#1.---}}}

\begin{document}
\title{Signatures of quark matter from neutron star mass-radius curve}

\author{Tuhin Malik} 
\email{tm@uc.pt}
\affiliation{Department of Physics, CFisUC, University of Coimbra, P-3004 - 516 Coimbra, Portugal}

\author{Milena Albino} 
\email{milena.albino@student.uc.pt}
\affiliation{Department of Physics, CFisUC, University of Coimbra, P-3004 - 516 Coimbra, Portugal}

\author{Pedro Sanson}
\email{pedro.sanson@student.uc.pt }
\affiliation{Department of Physics, CFisUC, University of Coimbra, P-3004 - 516 Coimbra, Portugal}

\author{Constança Providência}
\email{cp@uc.pt}
\affiliation{Department of Physics, CFisUC, University of Coimbra, P-3004 - 516 Coimbra, Portugal}

\begin{abstract}
Signatures of the presence of quark matter inside neutron stars are obtained from the analysis of more than 250,000 mass-radius curves that describe nucleonic, hyperonic and hybrid stars, based on five different hadronic relativistic mean-field models and four different descriptions of quark matter. A radius maximum occurring above $1.7\,M_\odot$ strongly favors quark matter inside the star. A second criterion identifying  the presence of quark matter is $X>-0.2$km, where  X=[R(2.0)-R(1.6)]-[R(1.6)-R(1.4)] is obtained from the radius of stars with 2.0, 1.6 and 1.4 $M_\odot$. It is also shown that most of the  hyperonic stars have a radius maximum that peaks around 1.5$M_\odot$ and compactness below 0.27 for the maximum-mass configuration, while for the nucleonic stars that have a radius maximum, this peaks below 1.3$M_\odot$ and the compactness of the maximum-mass configuration lies above 0.26.

\end{abstract}

\maketitle

\prlsec{Introduction} The determination  of the composition of neutron stars (NS) from their properties  is presently an active field of research. It is expected that these objects can provide information on the high density and low temperature region of the QCD phase diagram \cite{Watts:2014tja}, a region that is presently not accessible in the laboratory.  However,  it has been shown that recovering the proton fraction from the $\beta$-equilibrium equation of state (EOS) is challenging due to the degeneracy introduced through the isoscalar and isovector nuclear force channels \cite{Mondal:2021vzt,Imam:2021dbe,deTovar:2021sjo}. This could indicate that identifying non-nucleonic degrees of freedom inside NS may be even more complex, as different composition scenarios may produce similar behaviors of the EOS or the mass-radius NS curve. The inclusion of information  that raises the degeneracy will be necessary to distinguish different degrees of freedom.

Several agnostic descriptions have been used to identify the region in the pressure-energy density phase space that is populated by the current astrophysical and theoretical constraints and to determine the speed of sound behavior with density \cite{Tews:2018kmu,Annala:2019puf,Altiparmak:2022bke,Tan:2021ahl,annala2023strongly}. Even though these approaches are important for providing insight into the EOS covering  the unknown density range between the low density region constrained by chiral effective field theory \cite{Hebeler:2013nza,Drischler:2017wtt} and the high density constrained by perturbative QCD \cite{Kurkela:2009gj,Kurkela:2014vha,Komoltsev:2021jzg}, they do  not allow a clear determination of the NS composition and relevant degrees of freedom.

It has been shown that the  mass-radius  or mass-tidal deformability curves determined from a given EOS model may convey some information about the composition of the NS \cite{Han:2020adu,ferreira_2024,ferreira_2025,Cai:2025nxn,bauswein_2026,albino_2026}. The signatures proposed in these works rely on the knowledge of the slope and curvature of the mass-radius  or mass-tidal deformability curves. One expects that with the upcoming NS observatories, such as the enhanced X-ray Timing and Polarimetry mission—eXTP \cite{eXTP:2018anb}, Square Kilometer Array \cite{Watts:2014tja}, Einstein Telescope \cite{ET:2019dnz}, Cosmic Explorer \cite{reitze2019cosmic}, NewAthena  \cite{Cruise:2024mgo}, the uncertainties associated with the determination of the radius, mass, and tidal deformability will be much smaller.

In this letter, we show that the behavior of the mass-radius curve may provide a  clear indication of the NS composition, in particular, of the presence of quarks. It has been shown previously that a mass-radius curve with a negative slope for all  masses above one solar mass indicates that the underlying EOS contains, with a very large probability, only nucleonic degrees of freedom. If the mass-radius curve presents a range of masses with a positive slope, then there should be a radius maximum defined by $dR/dM=0$. In the present study, we  analyze the composition of stars that have a radius maximum  above one solar mass, and we show it allows us to distinguish between different compositions, in particular, that the mass of the maximum radius configuration gives a clear indication of the presence of quark degrees of freedom if it lies above 1.7 $M_\odot$. It is  also shown that the difference between the change of radius above and below a reference mass of 1.6 solar masses, i.e., the quantity $X=[R(2.0)-R(1.6)]-[R(1.6)-R(1.4)]$, identifies  the presence of quark matter in the core of the star if $X>-0.2$~km.
 We  complement this analysis by determining the correlation between the proton fraction and the location of $dR/dM=0$ mass and compactness, considering the nucleonic and hyperonic EOS. 
To carry out this study,  we consider more than 250,000 equations of state (EOS), including five datasets of nucleonic EOS \cite{Cartaxo:2025jpi}, five datasets of hyperonic EOS \cite{Sanson:2026lky}, and {four datasets of hybrid EOS} \cite{AlbinoLargeQData2025,AlbinoSmallQData2026}, constrained by nuclear matter properties, astrophysical observations, and theoretical constraints.

\prlsec{Methodology}
The present study is undertaken with fourteen sets of EOS obtained from RMF descriptions of baryonic matter  through Bayesian inference. The parameters of the models were obtained by imposing the following constraints: nuclear matter properties at saturation (binding energy, saturation density, incompressibility, and symmetry energy); the neutron EOS at low density from chiral effective field theory calculations; the perturbative QCD EOS at high densities through thermodynamic consistency; NICER results for the pulsars PSR J0030+0451, PSR J0437-4715, and PSR J0740+6620, as well as the tidal deformability from GW170817 by the LVC.\\
{\it Nucleonic EOS:} The nucleonic EOSs are discussed in \cite{Cartaxo:2025jpi}. They have been obtained with five different RMF models, four of which involve density dependent parameters (DDB, DDBy, DDMEX, DDMEXy), and one includes non-linear meson terms in the Lagrangian density and constant couplings (RMFNL)\\
{\it Hyperonic EOS:} The hyperonic EOSs are from \cite{Sanson:2026lky}. They have been obtained with the same models as the nucleonic EOS but include hyperons ($\Lambda$, $\Sigma$, and $\Xi$). The couplings of the hyperons to the scalar mesons are constrained by experimental data and to the isoscalar vector mesons, considering the SU(6) symmetry of the quark model. \\
{\it Hybrid EOS:} The hybrid EOSs have been obtained in \cite{AlbinoLargeQData2025} for hybrid stars with large quark cores and in \cite{AlbinoSmallQData2026} for hybrid stars with small quark cores. In the last case, quark matter is described by an NJL model with four and eight quark terms. For the latter, two different quark models are considered, the same NJL model as for small quark cores and the MFTQCD model, an MIT-like model with a repulsive vector term. In all cases, only configurations presenting a  radius maximum, i.e., $dR/dM=0$, above one solar mass are considered.  {For large quark cores, two NJL datasets released in Ref.~\cite{AlbinoLargeQData2025} are considered: NJL-GW (NJL) which was (not) inferred  with the GW170817 constraint. NJL and NJL-GW use the same quark model but different posterior constructions and prior ranges; their comparison therefore tests sensitivity to those choices.}

\begin{figure}[t]
\centering
\includegraphics[width=\linewidth]{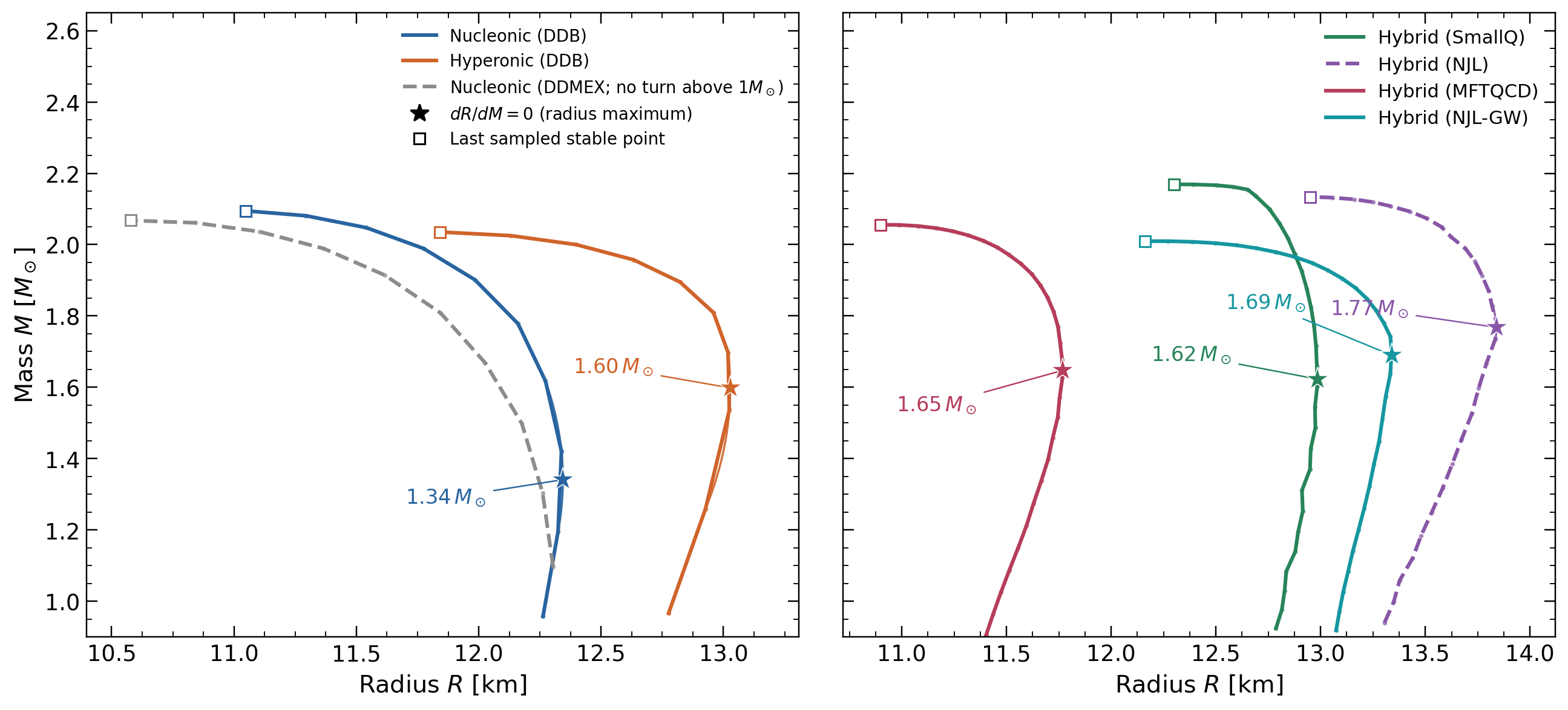}
\caption{Representative M--R curves: nucleonic and hyperonic EOS examples, including a no-turning-point control (left), {and SmallQ, NJL, MFTQCD and NJL-GW hybrid EOS (right).} Stars mark local radius maxima ($dR/dM=0$); labels give the corresponding stellar masses. Open squares mark the maximum mass.}
\label{fig:mr-turning-illustration}
\end{figure}

\prlsec{Results}
For illustrative purposes  of the radius maximum localization, we show in Fig.~\ref{fig:mr-turning-illustration} a selection of {seven EOS:} on the left panel, three nucleonic or hyperonic EOS;  on the right, {four hybrid EOS from different sets.}
The DDMEX curve is included to show an explicit example without a qualifying  radius maximum above $1\,M_\odot$. This is the only example with no qualifying radius maximum. The DDB nucleonic and hyperonic examples given are independent posterior draws that do not have the same parameters before and after adding hyperons. The radius maximum occurs at a smaller mass for the nucleonic EOS, a common feature of most nucleonic EOS.  The original examples were selected near the within-ensemble medians of their documented stellar summaries. The MFTQCD example was drawn uniformly at random with a fixed seed from EOSs with sampled two-solar-mass coverage and a strict interior radius maximum above $1\,M_\odot$.
The hybrid star mass-radius examples given show a radius maximum similar to that of hyperonic stars or greater. 

\begin{figure}[t]
\centering
\includegraphics[width=\columnwidth]{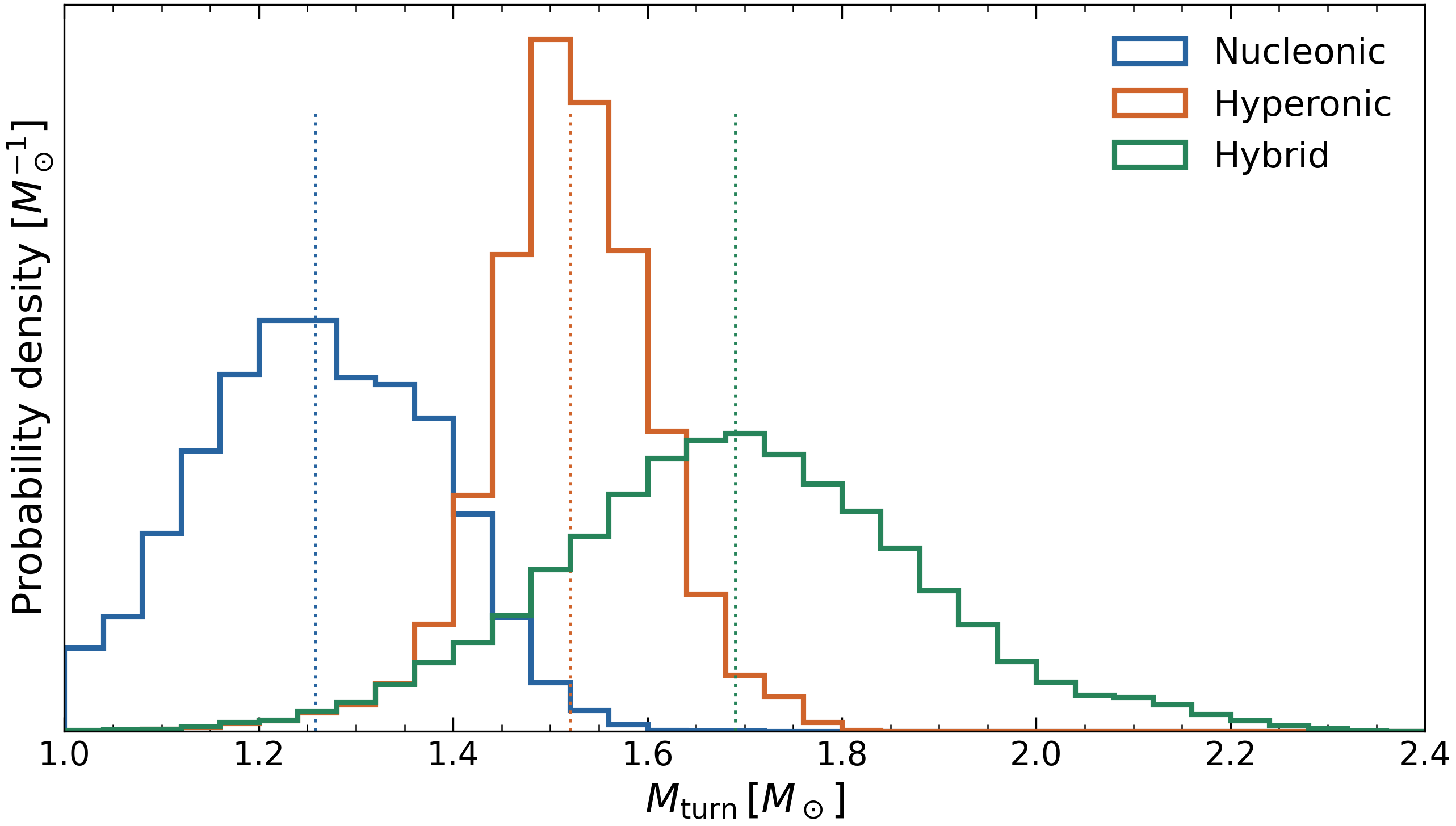}
\caption{Normalized distributions of the stellar mass at the maximum-radius configuration for nucleonic (blue), hyperonic (orange), and hybrid (green) EOSs. The hybrid distribution combines SmallQ, NJL, MFTQCD, and NJL-GW with equal weight for each dataset. Dotted lines indicate the medians.}
\label{fig:turning-mass-three-ensembles}
\end{figure}

\begin{figure*}[t]
\centering
\includegraphics[width=\textwidth]{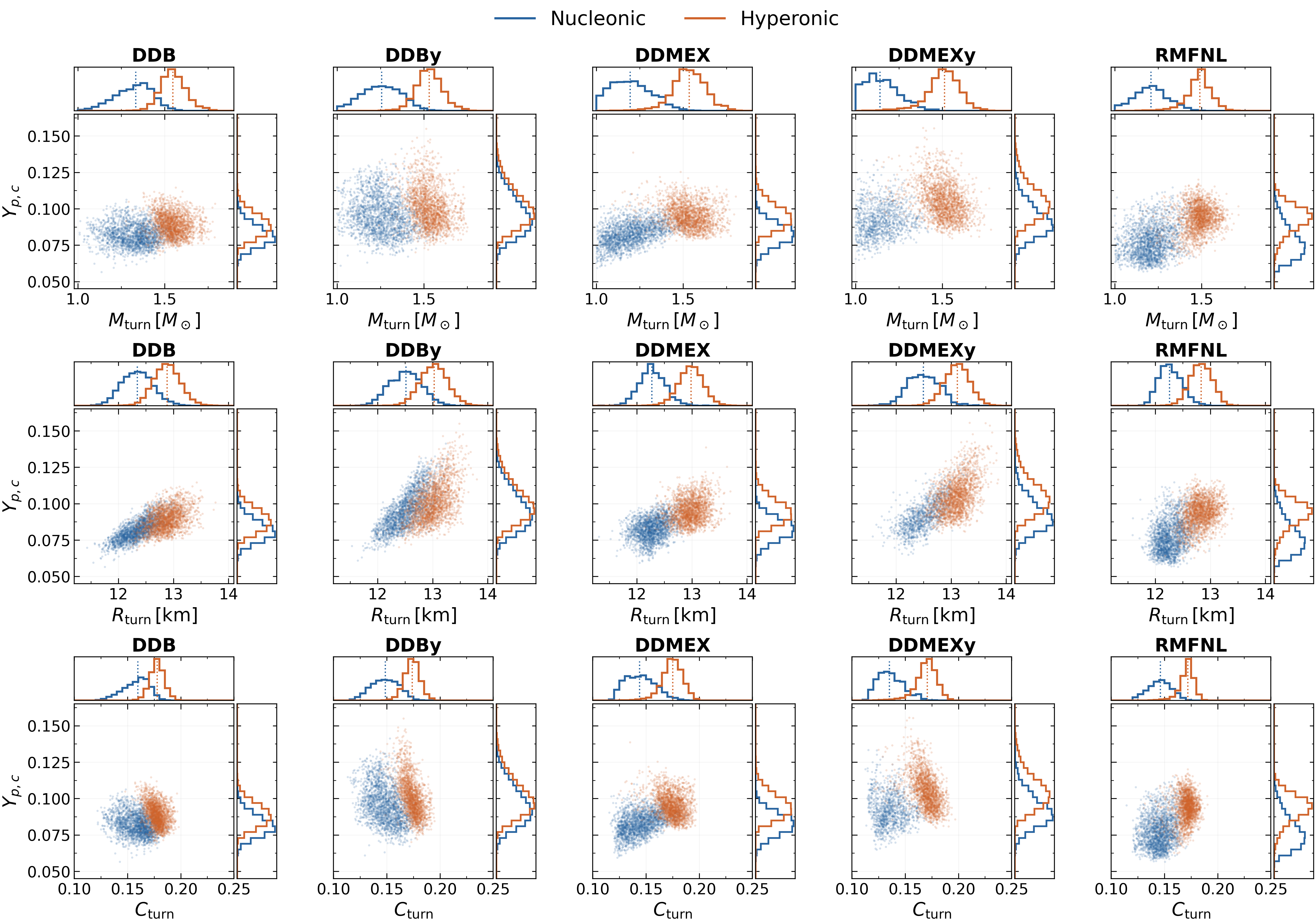}
\caption{Turning-point mass $M_{\rm turn}$ (top), radius $R_{\rm turn}$ (middle) and compactness $C_{\rm turn}$ (bottom) versus central proton fraction at the same star, for the five indicated EOS families. Blue/orange denote nucleonic/hyperonic ensembles. Scatter subsets and separately normalized full-sample marginals are shown; the dotted lines on the top histograms  mark horizontal-variable medians.}
\label{fig:turning-composition-families}
\end{figure*}

\begin{figure}[htb]
\centering
\includegraphics[width=\columnwidth]{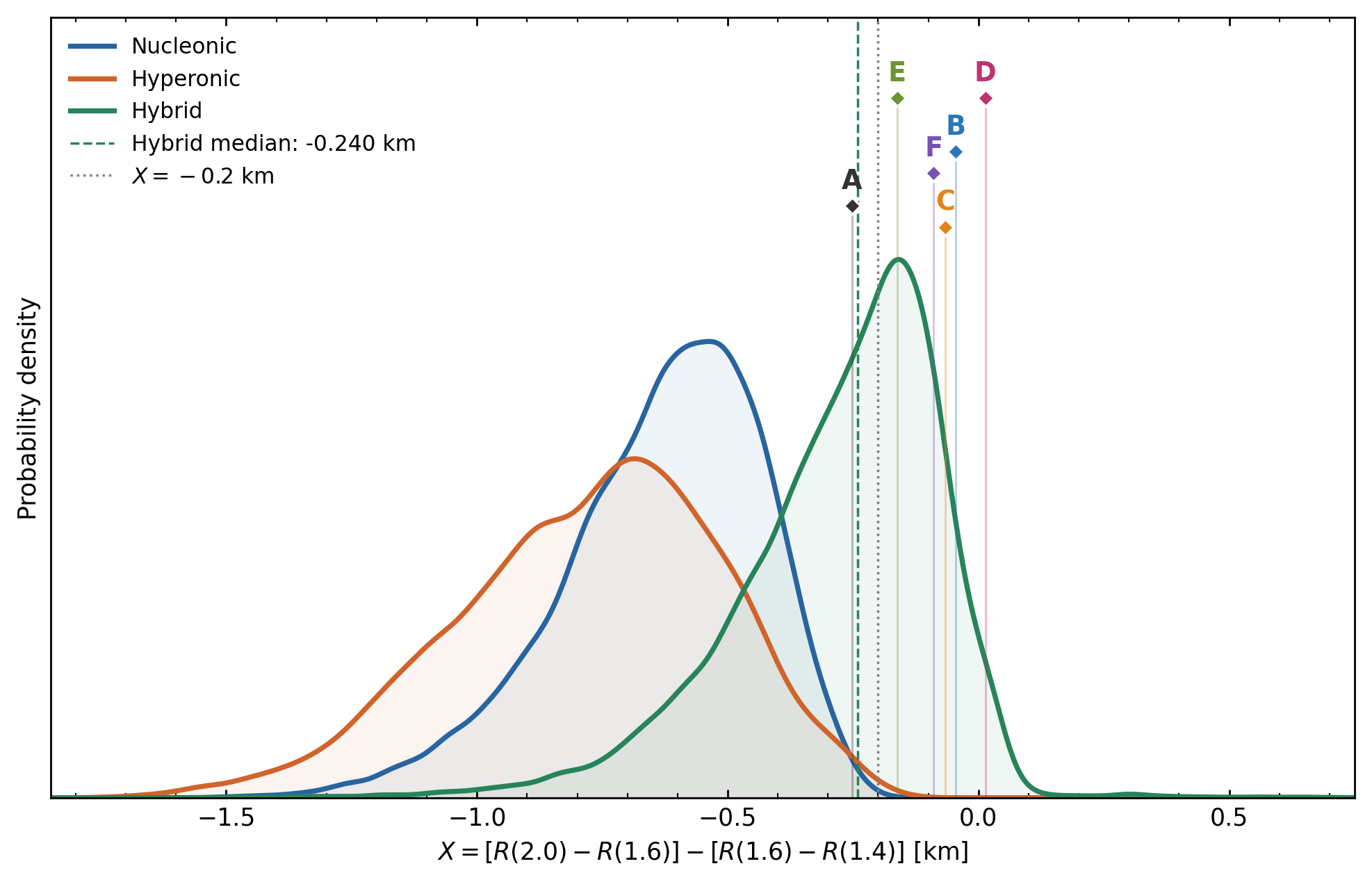}
\caption{Distributions of $X=[R(2.0)-R(1.6)]-[R(1.6)-R(1.4)]$ for nucleonic (blue), hyperonic (orange) and {pooled SmallQ+NJL+MFTQCD+NJL-GW hybrid (green) EOSs.} Categories have equal sample sizes, with {equal contributions from the four hybrid sets.} The dashed line marks the hybrid median, $-0.240\,\mathrm{km}$. {Diamonds A--F mark the six golden EOSs of Ref.~\cite{Ecker2025GoldenEOS}.  The gray dotted line marks $X=-0.2\,\mathrm{km}$.} 
}
\label{fig:full-catalog-radius-stiffness}
\end{figure}

We have  analyzed the location of the maximum radius  mass for all families of stars. In Fig. \ref{fig:turning-mass-three-ensembles}, we compare the normalized distributions of stellar mass at the radius maximum for nucleonic, hyperonic, and 
hybrid stars.
For the nucleonic and hyperonic histograms, the DDB, DDBy, DDMEX, DDMEXy and RMFNL groups have equal weight per selected EOS. The selection requires the highest-mass radius maximum above $1\,M_\odot$. Histograms are separately normalized with common $0.04\,M_\odot$ bins. {The hybrid curve pools SmallQ, NJL, MFTQCD and NJL-GW with equal total weight for each construction and equal weights within each construction. The selected samples contain 65,162 nucleonic, 125,773 hyperonic, and 24,396 hybrid EOSs (5,812 SmallQ, 6,512 NJL, 4,557 MFTQCD and 7,515 NJL-GW). Their class medians are 1.258, 1.521 and $1.691\,M_\odot$, respectively.} SmallQ uses Ref.~\cite{AlbinoSmallQData2026}; NJL, MFTQCD and NJL-GW use Ref.~\cite{AlbinoLargeQData2025}. While the distributions associated with the hybrid stars are quite broad, reflecting the model differences and the large and small quark cores imposed, they cover a range of high masses that neither the nucleonic nor the hyperonic stars do. This translates into the value of the  median of the different sets. The hyperonic distribution is the narrowest one, lying between the nucleonic and the hybrid distributions and showing a large overlap with the low mass tail of the hybrid distributions.  A distinct signature of a hybrid star seems to be the existence of the turning point at a high mass, above 1.7-1.8$M_\odot$.

The overall distribution of the  mass, radius,  and compactness of the stars at the turning point, i.e., at the radius maximum, is shown in Fig.~\ref{fig:turning-composition-families}. For each EOS, the selected turning point is the highest-mass with the radius maximum above $1\,M_\odot$, distinct from the maximum-mass configuration. In the figure, all three rows use the central proton fraction $Y_{p,c}=Y_p[n_c(M_{\rm turn})]$ of the turning point star. Compactness is defined as $C_{\rm turn}=GM_{\rm turn}/(R_{\rm turn}c^2)$. The  scatter panels use reproducible subsets of up to 2000 EOSs per ensemble. On the top and on the right side of each panel,  normalized marginal  histograms using the full selected ensembles are also shown. The posteriors for the nucleonic and hyperonic sets are independently inferred and unpaired. 
The main conclusions are: i) the peaks of the mass, radius, and compactness histograms at the turning point are distinct for the nucleonic stars in comparison with hyperonic stars across all  five sets, with hyperonic stars showing a peak at larger masses, radii, and compactnesses. In particular, the overlap of the turning point mass histograms is small, resulting in two almost independent distributions with medians at $\sim 1.25$ $M_\odot$ and $\sim1.5 \, M_\odot$. This property is reflected in the compactness, which also shows small overlapping distributions. For the hyperonic stars, the compactness peaks at $\sim 0.175$, while the compactness of nucleonic stars varies between $\sim 0.14$ and 0.16 depending on the underlying model. The proton fraction shows a moderate correlation with the turning point radius. 

\begin{figure}[t]
\centering
\includegraphics[width=\columnwidth]{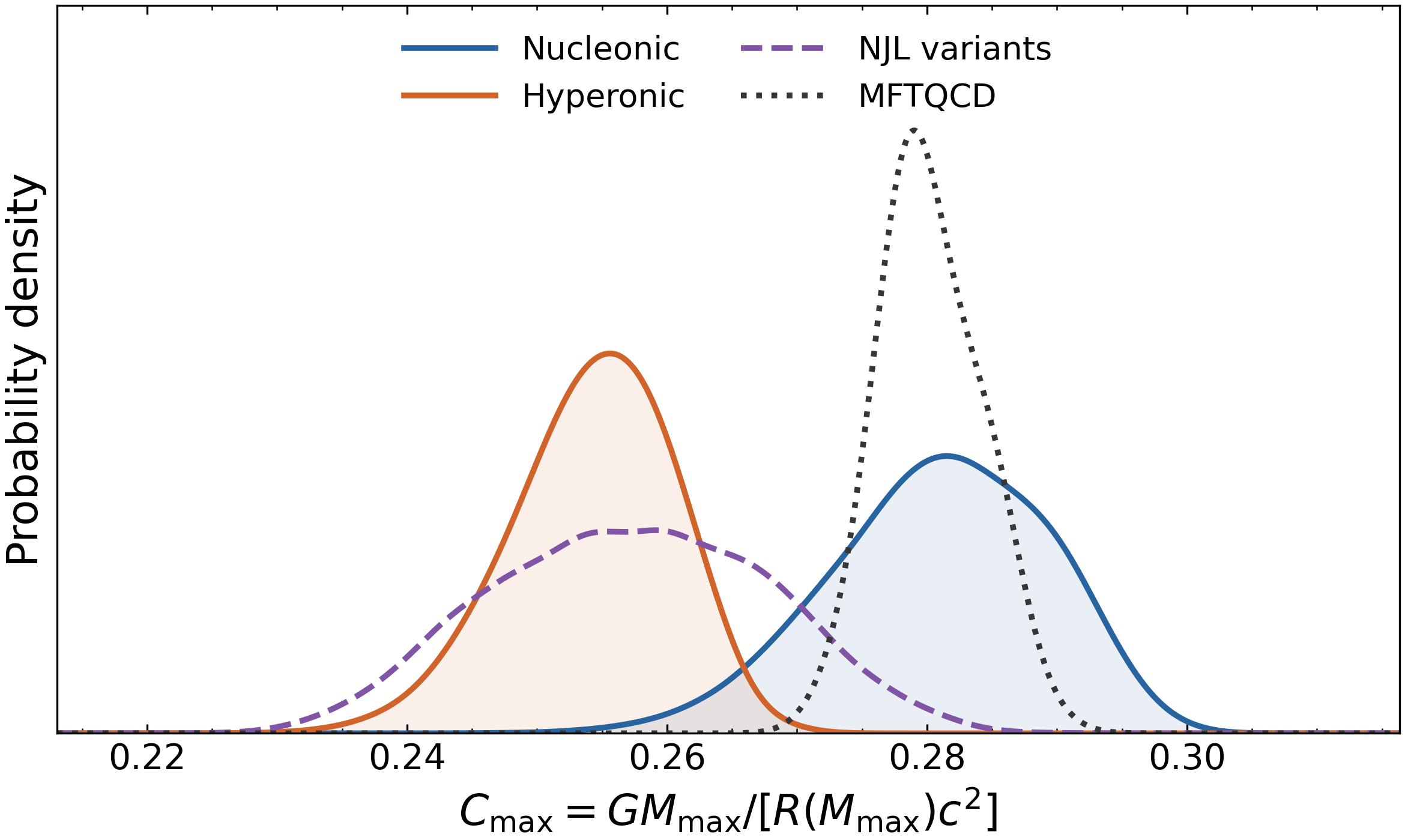}
\caption{Distributions of sampled maximum-mass compactness $C_{\max}=GM_{\max}/[R(M_{\max})c^2]$ for pooled nucleonic (blue), pooled hyperonic (orange), NJL-based hybrid (purple dashed), and MFTQCD hybrid (black dotted) EOSs. The NJL-based curve combines SmallQ, NJL and NJL-GW with equal total weight per construction. Each distribution is normalized separately; the same absolute kernel bandwidth is used throughout.}

\label{fig:pooled-cmax}
\end{figure}

Some  distinguishing features  of  the mass distribution of the maximum radius of  the hybrid stars with respect to the other two distributions are: there is only an overlap of its tail  with the tail of  nucleonic between  $\sim 1.2$ and $1.6\, M_\odot$;  it  overlaps below  1.6$M_\odot$ with the hyperonic distribution.  We have taken the radius of a 1.6$M_\odot$ star as a reference star and compared the change of radius between 1.4 and 1.6$M_\odot$ stars with the change of radius between 1.6 and 2.0$M_\odot$ stars. To compare the upper- and lower-interval radius changes, we introduce the observable
\begin{equation}
X=[R(2.0)-R(1.6)]-[R(1.6)-R(1.4)],
\label{eq:radius-contrast}
\end{equation}
where $R(a)$ denotes the radius of the $a\,M_\odot$ star for the same EOS; $X$ is measured in km.                                                                   
Fig.~\ref{fig:full-catalog-radius-stiffness} shows the class-conditional distributions of this observable.
The nucleonic and hyperonic groups pool the same five RMF families with per-EOS weighting. Each category uses a reproducible random sample of equal total size.
{Within the hybrid sample, SmallQ, NJL, MFTQCD and NJL-GW contribute equal numbers, using Refs.~\cite{AlbinoSmallQData2026,AlbinoLargeQData2025}.}  
In these sampled catalogs, the hybrid distribution is shifted toward larger $X$, with the median {$X=-0.240\,\mathrm{km}$}, while nucleonic and hyperonic predictions overlap substantially. However,  the hyperonic distribution spreads to smaller values, reflecting the fact that hyperonic stars go through a stronger radius decrease after hyperon onset.
{
With equal contributions from the four hybrid sets, approximately 41\% of the sampled hybrid EOSs have $X>-0.2\,\mathrm{km}$.
} 
The threshold $X_{\rm obs}>-0.2\,\mathrm{km}$ may therefore be explored as a phenomenological selection criterion within these EOS ensembles.

As an external comparison, we evaluate $X$ for the six golden EOSs A--F of Ecker et al.~\cite{Ecker2025GoldenEOS}, selected to span the central 68\% region of a theory- and observation-conditioned, nonparametric EOS ensemble. All six golden EOSs incorporate chiral-EFT and pQCD constraints by construction~\cite{Ecker2025GoldenEOS}. \footnote{See Table~S1 in the Supplemental Material for their properties.} Five (B--F) have a radius maximum at 1.7$M_\odot$ (E) or above (B-D,F) and have
$X>-0.2\,\mathrm{km}$, falling inside the regions favored by our hybrid catalogs;  their locations favor a hybrid-EOS interpretation relative to the nucleonic and hyperonic ensembles examined here, providing a strong phenomenological hint within this framework.
EOS A has  no radius maximum above 1$M_\odot$ and  has $X=-0.252\,\mathrm{km}$, both properties indicating that it probably contains no quark matter and could  represent a nucleonic EOS.

We have considered  119 archived cold CompOSE entries~\cite{ComposeAuditCompOSE2022}. After data curation: composition verification, duplicate removal,  imposition of  chEFT and pQCD constraints, we were left with  nine nucleonic, zero hyperonic and eight hybrid EOS. All nine nucleonic entries have $X<-0.4\,\mathrm{km}$; four of eight hybrid entries have $X>-0.20\,\mathrm{km}$. (see Figure~\ref{fig:compose-retained-rug}  and Supplemental Material for more details).

Figure~\ref{fig:pooled-cmax} combines the five families into a single  nucleonic and a single hyperonic distribution of the sampled maximum-mass compactness $C_{\max}=GM_{\max}/[R(M_{\max})c^2]$. The figure also includes  the three NJL hybrid families together in a single distribution, and the MFTQCD distribution.  For the nucleonic and hyperonic, the same eligible EOSs as in the family-resolved compactness--proton-fraction comparison are retained; that comparison is provided in the Supplemental Material. These two distributions were obtained with equal weight per selected EOS, the pooled medians are 0.2814 for nucleonic matter and 0.2547 for hyperonic matter. Our sets indicate that the compactness of the maximum-mass configuration is larger for nucleonic stars and spans between $\sim0.26$ and $\sim0.30$, while for hyperonic stars it lies between $\sim0.24$ and $\sim0.27$. The NJL-hybrid  distribution overlaps with the hyperonic one, showing a broader distribution. For these  EOS the deconfinement phase transtion occurs at $\gtrsim2n_0$.
 The MFTQCD-hybrid distribution, corresponding to EOS which predict a deconfinement phase transition at $\gtrsim n_0$, peaks on top of the nucleonic distribution, with a narrower distribution. The compactness alone will not allow to distinguish hybrid stars from nucleonic or hyperonic stars, only together with features as the radius maximum and the value of $X$. 

\prlsec{Conclusions}
We have analyzed the information that the mass-radius curves convey about the NS composition and, in particular, examined the properties of the turning point of the mass-radius curve corresponding to the maximum radius star, i.e., $dR/dM=0$. In addition, the correlation between the proton fraction and several NS properties was examined for nucleonic and hyperonic stars.

We show that the maximum radius configuration peaks at different masses for nucleonic and hyperonic stars. In addition, for the  hybrid star configurations, the high mass range of the distribution, for masses above 1.7-1.8$M_\odot$, was identified as exclusive to these stars. We propose the property $X=[R(2.0)-R(1.6)]-[R(1.6)-R(1.4)]$, as identifying hybrid stars when $X>-0.2$ km. 
These features were verified with CompOSE EOS that satisfy the same constraints our datasets.

The agnostic "golden" EOSs of Ecker et al.~\cite{Ecker2025GoldenEOS} serve as representative benchmarks of dense matter constrained by chiral EFT, pQCD, and astrophysical observations, yet their microscopic composition remains unspecified. For the first time, our analysis provides a composition-sensitive interpretation of these benchmarks: five of the six occupy the hybrid-dominated region identified by our analysis, offering a strong phenomenological indication of quark matter inside neutron stars. 

\prlsec{Acknowledgments} This work was partially supported by Portuguese national funds from FCT (Fundação para a Ciência e a Tecnologia, I.P., Portugal) under project 2024.16290.PEX with DOI identifier 10.54499/2024.16290.PEX and under project UID/04564/2025, identified by DOI 10.54499/UIDB/04564/2025. The authors acknowledge the Laboratory for Advanced Computing at the University of Coimbra for providing HPC resources that have contributed to the research results reported within this paper, URL: https://www.uc.pt/lca. The computational component of this research was also carried out on the Deucalion HPC platform in Portugal within the EURO HPC project EHPC-REG-2025R01-021, whose resources and technical assistance are gratefully acknowledged.

\clearpage
\onecolumngrid
\begin{center}\textbf{Supplemental Material}\end{center}
\renewcommand{\theHfigure}{S\arabic{figure}}
\renewcommand{\theHtable}{S\arabic{table}}

\renewcommand{\thefigure}{S\arabic{figure}}
\renewcommand{\thetable}{S\arabic{table}}
\setcounter{figure}{0}
\setcounter{table}{0}
\section{Family-resolved compactness and proton fraction}
\begin{figure}[!htb]
\centering
\includegraphics[width=\textwidth]{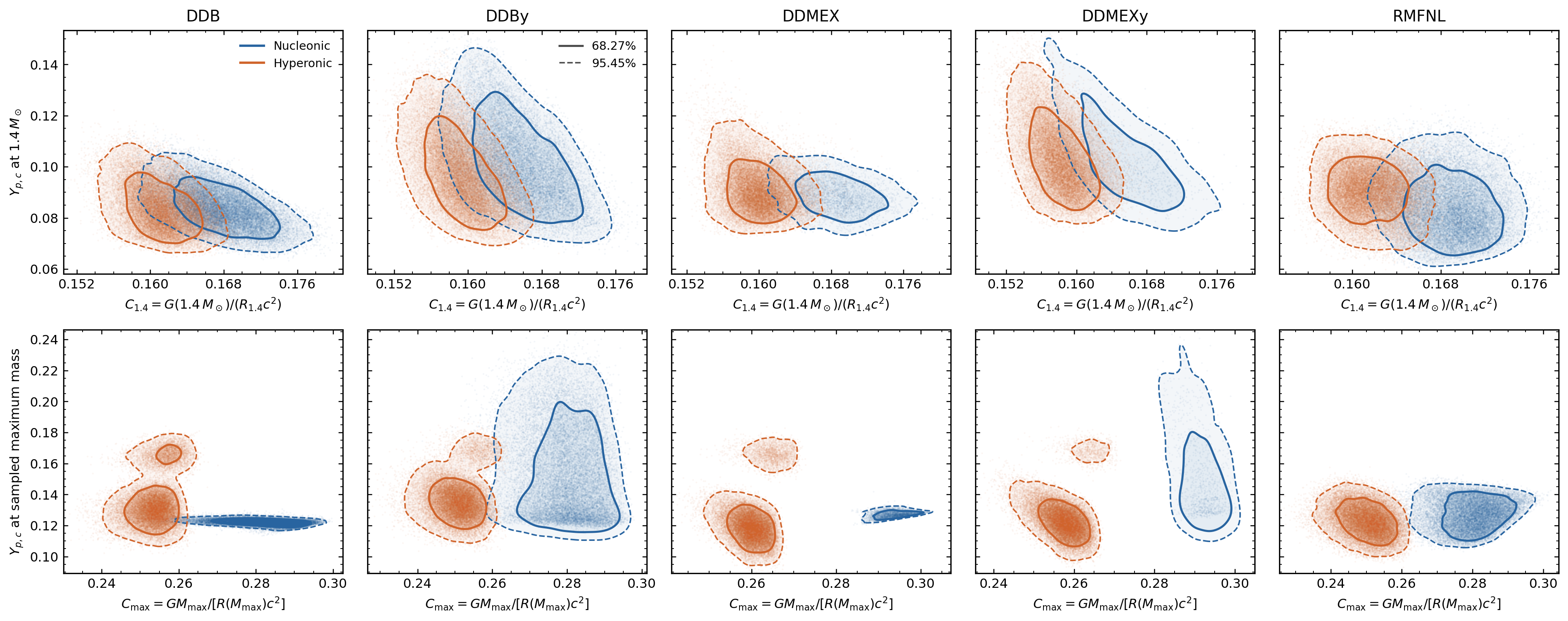}
\caption{Central proton fraction versus compactness $C=GM/(Rc^2)$ for the five indicated families. Rows show the $1.4\,M_\odot$ star (top) and sampled maximum-mass star (bottom), using the same eligible EOSs. The turning-point comparison appears in main-text Fig.~3. Blue/orange denote nucleonic/hyperonic ensembles. Faint points show samples; solid/dashed contours enclose 68.27\%/95.45\% of each model-ensemble density.
}
\label{fig:compactness-proton-fraction}
\end{figure}
Compactness $C=GM/(Rc^2)$ is a property that combines the mass and radius of the star, and in the following, we discuss the information it can convey when discussing composition.  In Fig.~\ref{fig:compactness-proton-fraction}, {we plot the central proton fraction as a function of compactness for $1.4\,M_\odot$ stars and at the sampled maximum mass, $C_{\max}=GM_{\max}/[R(M_{\max})c^2]$. The turning-point comparison is shown in main-text Fig.~3.} The proton fraction is evaluated at the star's central density. Both rows retain the original common EOS sample with valid mappings at all three configurations (including the turning-point star), conditional on a qualifying highest-mass bracket-supported radius maximum above $1\,M_\odot$ and a bracketed sampled mass maximum. Solid/dashed full-covariance Gaussian-kernel contours enclose 68.27\%/95.45\% of each ensemble density; these are model-ensemble highest-density regions. 

{The turning-point comparison in main-text Fig.~3 shows larger compactness for hyperonic stars due to their greater mass. In Fig.~\ref{fig:compactness-proton-fraction}, $1.4\,M_\odot$ hyperonic stars instead have smaller compactness than nucleonic stars because the former have larger radii.} In addition, they are indistinguishable concerning the proton fraction. In fact, 1.4$M_\odot$ hyperonic stars, if any, have only very small amounts of hyperons, so their compositional properties are defined by the proton fraction. The small compactness due to larger radii is caused by the greater stiffness of the EOS necessary to describe two solar mass stars. At the maximum mass, the compactness and proton fraction of hyperonic matter are  not much model dependent, in contrast to nucleonic matter, which shows a large dispersion in proton fraction, as already discussed in \cite{sup-Cartaxo:2025jpi}.  This reflects the fact that the hyperon-meson couplings have been constrained by experiments and that the onset of hyperons is favoured to reduce the  neutron Fermi pressure. Our sets indicate that the compactness of the maximum mass configuration is larger for nucleonic stars and  spreads between $\sim   0.26$  and  $\sim 0.3 $, while for hyperonic stars, it lies between $\sim$0.24 and $\sim$0.27.

\clearpage
\section{Golden equations of state}
\label{sec:golden-eos}
The six golden EOSs A--F of Ecker et al.~\cite{sup-Ecker2025GoldenEOS} were selected from a nonparametric posterior ensemble constrained by chiral EFT, pQCD and astrophysical observations. Table~\ref{tab:golden-turning-X} collects their radius-maximum masses and radii and the radius contrast $X=R(2.0)-2R(1.6)+R(1.4)$. EOS A has no resolved radius maximum above $1\,M_\odot$ on the published sequence.

\begin{table}[!htb]
\centering\small
\caption{Turning-point properties and radius contrast of the six golden EOSs of Ref.~\cite{sup-Ecker2025GoldenEOS}. $M_{\rm turn}$ and $R_{\rm turn}$ locate the highest-mass radius maximum above $1\,M_\odot$, estimated by a local quadratic fit to the published M--R grid. A dash indicates no resolved qualifying maximum (EOS A). $X$ uses linear interpolation at the three reference masses; tabulated digits describe grid estimates.}
\label{tab:golden-turning-X}
\begin{tabular}{lrrr}
\hline
EOS & $M_{\rm turn}$ & $R_{\rm turn}$ & $X$\\
& $[M_\odot]$ & $[\mathrm{km}]$ & $[\mathrm{km}]$\\
\hline
A & --- & --- & $-0.252$\\
B & 1.821 & 12.670 & $-0.045$\\
C & 1.853 & 12.873 & $-0.066$\\
D & 1.974 & 12.248 & $+0.014$\\
E & 1.689 & 12.589 & $-0.162$\\
F & 1.795 & 12.469 & $-0.089$\\
\hline
\end{tabular}
\end{table}

\section{CompOSE benchmark}
\label{app:compose-X}
From 119 archived cold CompOSE entries~\cite{sup-ComposeAuditCompOSE2022}, composition verification, duplicate removal, separation of Delta-resonance matter and radius coverage at all three reference masses select 90 entries: 65 nucleonic, six hyperonic and 19 hybrid. Their identifiers and references are listed in Table~\ref{tab:compose-models}.

We apply the same two filters to every class: containment of the tabulated beta-equilibrium pressures in the five adopted chiral-EFT intervals~\cite{sup-rutherford2024constraining} at $n=0.092,0.120,0.160,0.200,0.240\,\mathrm{fm}^{-3}$, and nonzero support for a causal, thermodynamically stable continuation from $n_L=1.20\,\mathrm{fm}^{-3}$ to the pQCD anchor $\mu_H=2.6\,\mathrm{GeV}$~\cite{sup-Komoltsev:2021jzg,sup-komoltsev_2023_7781233}, with log-uniform renormalization scale over $[1,4]$. Pressure and energy density are interpolated at the common $n_L$ without extrapolation. The pressure screen retains 19 nucleonic, zero hyperonic and eight hybrid entries; the subsequent pQCD test leaves nine, zero and eight, respectively. The pQCD test concerns the continuation above $n_L$; no separate internal-causality or maximum-mass stability filter is applied.

\begin{figure}[!htb]
\centering
\includegraphics[width=0.90\textwidth]{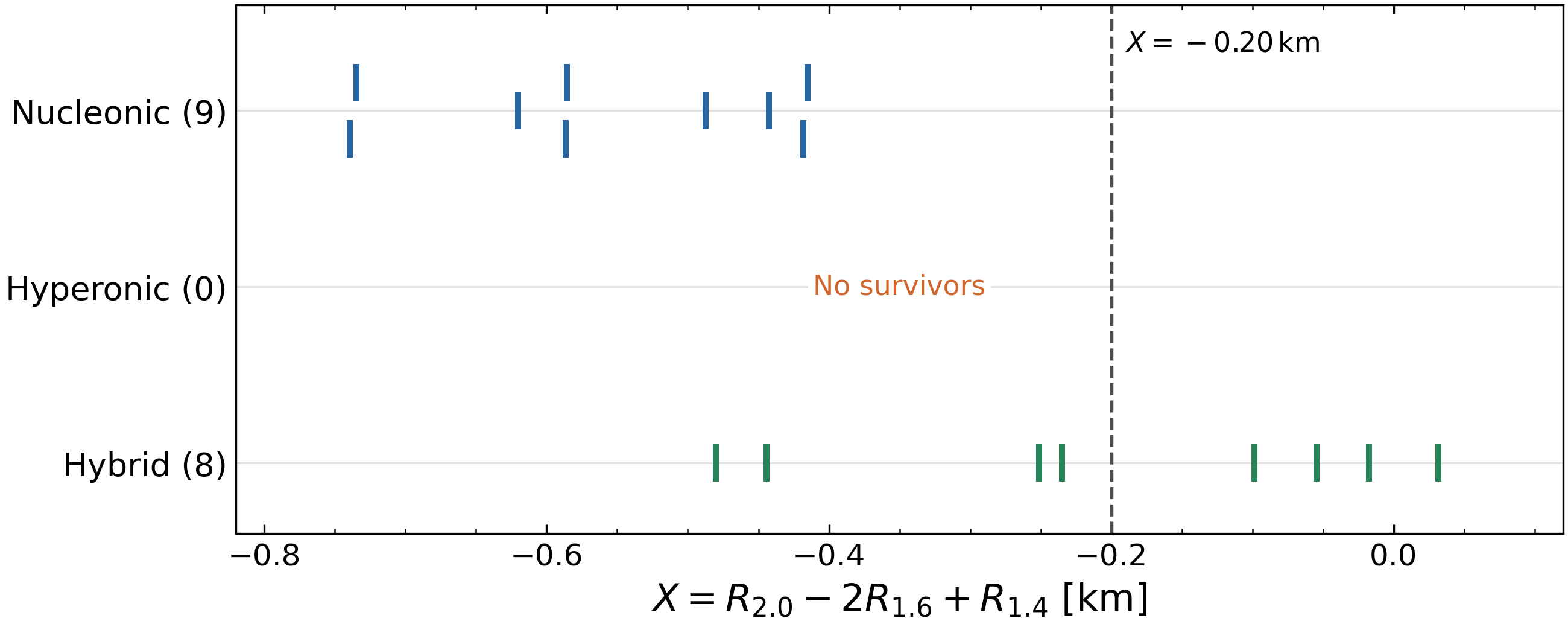}
\caption{Individual $X$ values of the CompOSE EOSs retained after the chiral-EFT beta-equilibrium pressure and pQCD connection filters at $n_L=1.20\,\mathrm{fm}^{-3}$. Each tick represents one EOS from Table~\ref{tab:compose-survivors}; nearby ticks are staggered vertically for visibility. Nine nucleonic and eight hybrid entries survive, with no hyperonic survivors. The dashed vertical line marks $X=-0.20\,\mathrm{km}$. No density smoothing is applied; the pressure-convention qualifications in the CompOSE comparison also apply here.}
\label{fig:compose-retained-rug}
\end{figure}

Figure~\ref{fig:compose-retained-rug} shows only these retained EOSs, listed in Table~\ref{tab:compose-survivors}. All nine nucleonic entries have $X<-0.24\,\mathrm{km}$; four of eight hybrid entries have $X>-0.20\,\mathrm{km}$. QMC-RMF3 is electron-only whereas the pressure reference includes muons; QHC18's hadronic lepton convention remains unverified. Their numerical passes are marked in the table; excluding them leaves eight nucleonic and seven hybrid entries with matched or verified conventions. Hybrid EOS labels do not by themselves establish quark cores at the reference masses.

\begin{table}[!htb]
\centering\small
\caption{EOSs retained in Fig.~\ref{fig:compose-retained-rug} after the chiral-EFT pressure and pQCD filters at $n_L=1.20\,\mathrm{fm}^{-3}$. Daggers mark the pressure-convention qualifications discussed in the text.}
\label{tab:compose-survivors}
\begin{tabular}{lrl}
\hline
Class & Number & Retained models\\
\hline
Nucleonic &9&\parbox[t]{0.70\textwidth}{\raggedright GMSR(SLy5), QMC-RMF3$^{\dagger}$, RG(SkI4), RG(SkI6), RG(SLy2), RG(SLy4), RG(SLy9), GPPVA(TW), XMLSLZ(TW99)}\\[5pt]
Hyperonic &0&None\\[3pt]
Hybrid &8&\parbox[t]{0.70\textwidth}{\raggedright QHC18$^{\dagger}$, QHC21$\chi$ A, B, C, D; V-QCD(APR) soft, intermediate, stiff}\\[3pt]
\hline
\end{tabular}
\end{table}

\begin{table}[!htb]
\centering\small
\caption{All 90 plotted CompOSE entries, grouped by release/construction with family references. Listed identifiers distinguish the archived table variants; their weights are per entry. CMF1/3/5 are hyperonic and CMF2/4/6 nucleonic. Composition classes follow documented particle content, not interaction names: the GPPVA(FSU2H) table used here is nucleonic.}
\label{tab:compose-models}
\begin{tabular}{lll}
\hline
\parbox[t]{0.19\textwidth}{\raggedright Family} & \parbox[t]{0.61\textwidth}{\raggedright Archived table identifiers} & \parbox[t]{0.09\textwidth}{\raggedright Sources}\\[3pt]
\hline
\parbox[t]{0.19\textwidth}{\raggedright BL chiral} & \parbox[t]{0.61\textwidth}{\raggedright BLchiral, BLchiral\_unified} & \parbox[t]{0.09\textwidth}{\raggedright \cite{sup-ComposeAuditBL2018}}\\[3pt]
\parbox[t]{0.19\textwidth}{\raggedright APR} & \parbox[t]{0.61\textwidth}{\raggedright aprapr} & \parbox[t]{0.09\textwidth}{\raggedright \cite{sup-ComposeAuditAPR1998}}\\[3pt]
\parbox[t]{0.19\textwidth}{\raggedright CMF hadronic} & \parbox[t]{0.61\textwidth}{\raggedright cmf1, cmf2, cmf3, cmf4, cmf5, cmf6} & \parbox[t]{0.09\textwidth}{\raggedright \cite{sup-ComposeAuditCMF2021,sup-ComposeAuditCMFHybrid2022}}\\[3pt]
\parbox[t]{0.19\textwidth}{\raggedright CMF hybrid} & \parbox[t]{0.61\textwidth}{\raggedright cmf5hybrid, cmf6hybrid} & \parbox[t]{0.09\textwidth}{\raggedright \cite{sup-ComposeAuditCMFHybrid2022}}\\[3pt]
\parbox[t]{0.19\textwidth}{\raggedright Gogny D1M*} & \parbox[t]{0.61\textwidth}{\raggedright d1mstar} & \parbox[t]{0.09\textwidth}{\raggedright \cite{sup-ComposeAuditD1Mstar2018}}\\[3pt]
\parbox[t]{0.19\textwidth}{\raggedright GPPVA unified RMF} & \parbox[t]{0.61\textwidth}{\raggedright dd2, ddme2, fsu2, fsu2h, fsu2r, nl3wrl55, tm1e, tw} & \parbox[t]{0.09\textwidth}{\raggedright \cite{sup-ComposeAuditGPPVA2014,sup-ComposeAuditBSk2018}}\\[3pt]
\parbox[t]{0.19\textwidth}{\raggedright CMGO GDFM-I} & \parbox[t]{0.61\textwidth}{\raggedright gdfm1} & \parbox[t]{0.09\textwidth}{\raggedright \cite{sup-ComposeAuditCMGO2023}}\\[3pt]
\parbox[t]{0.19\textwidth}{\raggedright GMSR} & \parbox[t]{0.61\textwidth}{\raggedright gmsrdhsl59, gmsrdhsl69, gmsrh1, gmsrh2, gmsrh3, gmsrh4, gmsrh5, gmsrh7, gmsrsly5} & \parbox[t]{0.09\textwidth}{\raggedright \cite{sup-ComposeAuditGMSR2022}}\\[3pt]
\parbox[t]{0.19\textwidth}{\raggedright PCGS PCSB} & \parbox[t]{0.61\textwidth}{\raggedright pcgsb0, pcgsb1, pcgsb2} & \parbox[t]{0.09\textwidth}{\raggedright \cite{sup-ComposeAuditPCSB2023}}\\[3pt]
\parbox[t]{0.19\textwidth}{\raggedright PCP BSk} & \parbox[t]{0.61\textwidth}{\raggedright pcpbsk22, pcpbsk24, pcpbsk25, pcpbsk26} & \parbox[t]{0.09\textwidth}{\raggedright \cite{sup-ComposeAuditBSk2018}}\\[3pt]
\parbox[t]{0.19\textwidth}{\raggedright QMC-RMF} & \parbox[t]{0.61\textwidth}{\raggedright qmcrmf2, qmcrmf3, qmcrmf4} & \parbox[t]{0.09\textwidth}{\raggedright \cite{sup-ComposeAuditQMCRMF2022}}\\[3pt]
\parbox[t]{0.19\textwidth}{\raggedright RG Skyrme} & \parbox[t]{0.61\textwidth}{\raggedright rgrs, rgsk255, rgsk272, rgska, rgskb, rgski2, rgski3, rgski4, rgski5, rgski6, rgskmp, rgsly2, rgsly230a, rgsly4, rgsly9} & \parbox[t]{0.09\textwidth}{\raggedright \cite{sup-ComposeAuditRG2015}}\\[3pt]
\parbox[t]{0.19\textwidth}{\raggedright SPG} & \parbox[t]{0.61\textwidth}{\raggedright spgm1, spgm2, spgm3, spgm4, spgm5} & \parbox[t]{0.09\textwidth}{\raggedright \cite{sup-ComposeAuditSPG}}\\[3pt]
\parbox[t]{0.19\textwidth}{\raggedright XMLSLZ} & \parbox[t]{0.61\textwidth}{\raggedright xmslzddlz1, xmslzddme2, xmslzddmex, xmslzgm1, xmslzmtvtc, xmslznl3, xmslzpk1, xmslzpkdd, xmslztm1, xmslztw99} & \parbox[t]{0.09\textwidth}{\raggedright \cite{sup-ComposeAuditXMLSLZ2022}}\\[3pt]
\parbox[t]{0.19\textwidth}{\raggedright DD2-FRG} & \parbox[t]{0.61\textwidth}{\raggedright dd2frg2, dd2frg2vecmes} & \parbox[t]{0.09\textwidth}{\raggedright \cite{sup-ComposeAuditDD2FRG2020}}\\[3pt]
\parbox[t]{0.19\textwidth}{\raggedright QHC18} & \parbox[t]{0.61\textwidth}{\raggedright qhc18} & \parbox[t]{0.09\textwidth}{\raggedright \cite{sup-ComposeAuditQHC2018}}\\[3pt]
\parbox[t]{0.19\textwidth}{\raggedright QHC19} & \parbox[t]{0.61\textwidth}{\raggedright qhc19, qhc19c, qhc19d} & \parbox[t]{0.09\textwidth}{\raggedright \cite{sup-ComposeAuditQHC2019}}\\[3pt]
\parbox[t]{0.19\textwidth}{\raggedright QHC21} & \parbox[t]{0.61\textwidth}{\raggedright qhc21T\_a, qhc21T\_b, qhc21T\_c, qhc21T\_d, qhc21chi\_a, qhc21chi\_b, qhc21chi\_c, qhc21chi\_d} & \parbox[t]{0.09\textwidth}{\raggedright \cite{sup-ComposeAuditQHC2021}}\\[3pt]
\parbox[t]{0.19\textwidth}{\raggedright V-QCD} & \parbox[t]{0.61\textwidth}{\raggedright vqcd\_intermediate, vqcd\_soft, vqcd\_stiff} & \parbox[t]{0.09\textwidth}{\raggedright \cite{sup-ComposeAuditVQCD2021}}\\[3pt]
\parbox[t]{0.19\textwidth}{\raggedright OPGR hyperonic} & \parbox[t]{0.61\textwidth}{\raggedright ddhdeltay4, gm1y5, gm1y6} & \parbox[t]{0.09\textwidth}{\raggedright \cite{sup-ComposeAuditOPGR2015}}\\[3pt]
\hline
\end{tabular}
\end{table}
\clearpage
\begingroup
\let\arxivOriginalLabel\label
\renewcommand{\label}[1]{\def\arxivLabelArg{#1}\def\arxivLastBib{LastBibItem}\ifx\arxivLabelArg\arxivLastBib\arxivOriginalLabel{SupplementLastBibItem}\else\arxivOriginalLabel{#1}\fi}

\endgroup%

\end{document}